\documentclass[a4paper, 11pt]{article} 

\usepackage{graphicx}
\usepackage{wrapfig}
\usepackage{lscape}
\usepackage{rotating}
\usepackage{epstopdf}
\usepackage{authblk}
\usepackage{units}
\usepackage{amsmath}
\usepackage[colorinlistoftodos]{todonotes}

\usepackage{setspace} 
\usepackage[margin=2.0cm]{geometry}
\usepackage{multicol}
\usepackage{siunitx}

\usepackage{subfig}
\usepackage{float}
\usepackage[skip=2pt, font=small]{caption}

\usepackage[colorlinks=true,linkcolor=blue,urlcolor=blue,citecolor=blue]{hyperref}
\usepackage[sort&compress,super,comma]{natbib}
\usepackage{url,doi}

\usepackage{soul}
\usepackage{textcomp}
\usepackage[right]{lineno}
\usepackage[symbol]{footmisc}

\begin{document}

\setlength\abovedisplayskip{2pt}
\setlength\belowdisplayskip{3pt}

\title{Enhancing surface production of negative ions using nitrogen doped diamond in a deuterium plasma}

\author[1$\dagger$]{Gregory J. Smith}
\author[1]{James Ellis\thanks{Currently at: Leibniz Institute for Plasma Science and Technology (INP), Felix-Hausdorff-Str. 2, 17489 Greifswald, Germany}}
\author[2]{Roba Moussaoui}
\author[2]{C\'edric Pardanaud}
\author[2]{Celine Martin}
\author[3]{Jocelyn Achard}
\author[3]{Riadh Issaoui}
\author[1]{Timo Gans}
\author[1]{James P. Dedrick}
\author[2]{Gilles Cartry}

\affil[1]{York Plasma Institute, Department of Physics, University of York, Heslington, York, YO10 5DD, UK}
\affil[2]{Aix-Marseille Universit\'e / CNRS, PIIM, UMR 6633, Centre de St J\'er\^ome, case 241, 13397 Marseille Cedex 20, , France}
\affil[3]{LSPM-CNRS, UPR 3407, Universit\'e Sorbonne Paris Nord Cit\'e, 99, avenue JB Cl\'ement, 93430 Villetaneuse, France}
\affil[$\dagger$]{E-mail: \href{mailto:gjs507@york.ac.uk}{gjs507@york.ac.uk} }

\maketitle

\begin{abstract}

The production of negative ions is of significant interest for applications including mass spectrometry, particle acceleration, material surface processing, and neutral beam injection for magnetic confinement fusion. Methods to improve the efficiency of the surface production of negative ions, without the use of low work function metals, are of interest for mitigating the complex engineering challenges these materials introduce. In this study we investigate the production of negative ions by doping diamond with nitrogen. Negatively biased ($-20$~V or $-130$~V), nitrogen doped micro-crystalline diamond films are introduced to a low pressure deuterium plasma (helicon source operated in capacitive mode, 2 Pa, 26 W) and negative ion energy distribution functions (NIEDFs) are measured via mass spectrometry with respect to the surface temperature (30$^{\circ}$C to 750$^{\circ}$C) and dopant concentration. The results suggest that nitrogen doping has little influence on the yield when the sample is biased at $-130$~V, but when a relatively small bias voltage of $-20$~V is applied the yield is increased by a factor of 2 above that of un-doped diamond when its temperature reaches 550$^{\circ}$C. The doping of diamond with nitrogen is a new method for controlling the surface production of negative ions, which continues to be of significant interest for a wide variety of practical applications.

\end{abstract}
\doublespacing

\section{\large{Introduction}}

The development of negative ion sources is of significant interest due to their applications in particle acceleration \cite{Ueno2010, Peters2000, Moehs2005, Lettry2014, Faircloth2018}, neutron generation  \cite{Welton2016, Antolak2016}, mass spectrometry \cite{Alton1994, Middleton1974, Calcagnile2005, Yoneda2004}, spacecraft propulsion \cite{Rafalskyi2016, Lafleur2015a, Aanesland2015}, nano-electronics manufacturing  \cite{Vozniy2009a}, and neutral beam heating for magnetic confinement fusion (MCF) \cite{Hemsworth2017, Hemsworth2009, Hemsworth2005, Fantz20127}.

One application of particular interest is the creation of negative-ion beams suitable for MCF neutral beam injection, which has a proposed requirement of accelerating a 40~A current of deuterium negative ions to 1~MeV~\cite{Hemsworth2017}. This primarily utilises negative ion surface production, as distinct from volume production, to increase the density of negative ions close to the extraction grid~\cite{Belchenko1974,Dudnikov2019, Dudnikov2019a}.

Negative ion production from plasma facing surfaces can be enhanced through the application of a low work function alkali metal~\cite{Dudnikov2019a}. Current methods apply a thin layer of caesium to the extraction region of the ion source \cite{Heinemann2009}. This is achieved by injecting caesium vapour into the plasma and allowing it to condense onto the inside of the ion source \cite{Tsumori2016}. There exist some limitations with this approach, such as controlling the application of the caesium so that it condenses in the right locations and at a rate that is sufficient to maintain an optimum thickness at the extraction grid \cite{Froschle2009}. Additionally, this method introduces complex engineering challenges, eg. equipment maintenance and potential for caesium pollution \cite{Cartry12017, Kurutz2017}. Alternative materials to caesium are therefore of interest. 

Several studies have been carried out to investigate negative ion production using alternative materials to caesium via their exposure to low pressure electronegative plasmas. Such materials include non-dielectric and dielectric materials including: diamond-like-carbon (DLC)\cite{Ahmad2014}, novel electrides \cite{Sasao}, highly orientated pyrolitic graphite (HOPG)\cite{Ahmad2014, Schiesko2009, Schiesko2010c, Cartry12017},  diamond \cite{Ahmad2014, Kumar2011, Schiesko2009, Schiesko2010c, Kogut2019, Achkasov2019, Ahmad2014, Kogut2017a, Dubois2016}, and low work function materials other than caesium (LaB\textsubscript{6}, MoLa)\cite{Kurutz2017}. 

Dielectric materials are of particular interest as an alternative to low work function metals \cite{Cartry12017}. Generally, for atoms approaching a surface, the affinity level of the atom is gradually downshifted until it overlaps with the surface material's valence band. Electrons can then tunnel from the valence band of the surface to the approaching atom and form a negative ion, this is the so-called resonant charge transfer (RCT) process, as summarised in Ref.~\citenum{Borisov2000}. For a metal, the conduction band is situated on top of the valence band. When a newly created ion begins to leave the surface, the probability of electron loss through tunnelling back to the conduction band of the surface is high due to the resonance between the affinity level of the negative ion and the empty states of the conduction band. This means that most metals produce negligible negative ions through surface ionisation processes \cite{Brako1989}. Unlike most metals, caesium can be used to enhance negative ion production because it has a low work function. This increases the distance at which the resonance between the affinity level of the new ion and the empty conduction states occurs, reducing the probability that the electron tunnels back to the surface \cite{Borisov2000, Brako1989}.

In contrast to metal surfaces, where the conduction band lies on top of the valence band, the band gap of dielectrics suppresses the tunnelling of electrons from a newly created negative ion back to the material's surface. This means that a new negative ion can travel a larger distance away from the surface before reaching a point where its affinity level is in resonance with the empty states of the conduction band. The increased distance of the ion from the surface reduces the probability that the electron associated with the new negative ion will tunnel into the empty states of the conduction band, thereby increasing the negative ion yield \cite{Cartry12017, J.Los1990}.

One potential drawback to the use of dielectric surfaces is that to generate a negative ion the atom-surface distance for a dielectric must be much smaller than for a metal. This is due to the larger energy gap that the occupied valence band states lie beneath the vacuum level \cite{Borisov2000}. Fortunately, the atom-surface interaction process is amplified by the Coulomb interaction between a negative ion and a localised hole in the surface material \cite{Borisov1996}. This can result in a high ionisation efficiency as demonstrated in beam experiments \cite{Winter2002, Roncin2002, Auth1998}. For these reasons, dielectrics are of interest as an alternative to low work function metals for the surface production of negative ions.

Carbon surfaces are one prospective category of materials that are of interest for replacing low work function metals where negative ions are to be produced. For instance, DLC has been used to produce negative ions from incoming neutral particles for a spacecraft particle detector, when low work function metals would not have been appropriate \cite{McComas2009}. Of the forms of carbon, diamond has particularly beneficial properties: 

\begin{itemize}
    
    \item It is a dielectric with a large band gap (5.5~eV)\cite{Diederich1998} that suppresses the destruction of negative ions as they leave the material's surface
    
    \item It can be grown to have `designer' properties such as the preferential growth of a particular crystal face to alter the electronic structure of its surface\cite{Diederich1998}

    \item When it is being grown, dopants can be introduced to change its effective work function and electron affinity \cite{Liu2017a, Tachibana2001, Scholze1996, Baranauskas1999}

    \item It can have a negative electron affinity when the surface is hydrogen terminated \cite{Diederich1998}, which reduces its effective work function by reducing the energy gap between the valence band and the vacuum level. This is thought to have a positive influence on negative ion production \cite{Cartry12017}

    \item When heated to 450$^{\circ}$C, diamond has previously been shown to produce five times more negative ions compared to other forms of carbon e.g. graphite \cite{Kumar2011}
    
\end{itemize}

As a means of increasing the production of negative ions, previous work with diamond has investigated using single, nano- and micro-crystalline diamond and also \textit{p}-type doping of micro-crystalline diamond (MCD) using boron \cite{Ahmad2014}. The \textit{n}-type doping of diamond using nitrogen has not previously been studied in this context and it is thought that it could lead to favourable properties for negative ion production for two reasons. Firstly, previous studies of the electronic properties of nitrogen doped diamond have demonstrated that nitrogen doping creates a deep donor level in the band gap of the diamond at 1.7~eV \cite{Yiming2014}. This lowers the effective work function to approximately 3.1~eV \cite{Diederich1999}, which is lower than boron doped diamond (3.9~eV) \cite{Diederich1999} and un-doped diamond ($\sim$4.5~eV, with hydrogenated surface and negative electron affinity) \cite{Abbott2001}. Secondly, it is thought that having the aforementioned deep donor level of electrons close to the vacuum level could increase the negative ion production from diamond by creating a source of electrons close to the vacuum level \cite{Bacal2015a}.

In this study, we investigate the production of negative ions from nitrogen doped diamond films in a low pressure deuterium plasma. Comparing micro-crystalline nitrogen doped diamond (MCNDD) with un-doped micro crystalline doped diamond (MCD) and previously investigated micro-crystalline boron doped diamond (MCBDD) \cite{Cartry12017, Kogut2019}, we consider `low energy' (11~eV) and `high energy' (48~eV) ion bombardment conditions at the surface as a mechanism for increasing the negative ion yield. The experimental methods are described in section~\ref{sec:expmeth}: plasma source in~\ref{sec:plasmasource}, sample holder in~\ref{sec:sampleholder} and the measurement method in~\ref{sec:Measurementandmassspec}. The micro-crystalline diamond samples are described in section~\ref{sec:Sampleprep}, with the surface characterisation using confocal microscopy and Raman spectroscopy described in~\ref{sec:surfacestudy}. The results are presented in section~\ref{sec:results}.

\section{\large{Method}}\label{sec:expmeth}

The experimental setup is shown in figure~\ref{Experiment}. It consists of a low pressure deuterium plasma source, a temperature controlled sample holder, and a mass spectrometer for the measurement of negative ions produced at the diamond film's surface.

\subsection{\normalsize{Description of the plasma source}}\label{sec:plasmasource}

A deuterium plasma, figure~\ref{Experiment}~(a), is produced via a helicon source operated in capacitive mode (2~Pa, 26~W), which then expands into a diffusion chamber\cite{AAhmad12013}. The pressure of the diffusion chamber, as measured by a Baratron gauge (MKS), is regulated via a mass flow controller (7.6~sccm, BROOKS 5850TR) in combination with a 150~mm diameter Riber gate valve installed in front of a turbo molecular pump (Alcatel ATP400). To reduce experimental drifts, the experiment source chamber and lower spherical diffusion chamber have a base pressure of 10\textsuperscript{-5}~Pa, which is lower than the base pressure of a previous setup of 10\textsuperscript{-4}~Pa \cite{AAhmad12013}. 

The relatively low power coupled to the plasma source results in plasma densities of approximately 10\textsuperscript{14}~m\textsuperscript{-3} in the spherical diffusion chamber \cite{AAhmad12013}. The choice of power and pressure was for similarity with previous work\cite{Achkasov2019}. The positive ion composition of the deuterium plasma is measured by the mass spectrometer, described below, to be (84~$\pm$~2)~\% D\textsubscript3\textsuperscript{+} ions, (14~$\pm$~2)~\% D\textsubscript2\textsuperscript{+} ions and (1.1~$\pm$~0.2)~\% D\textsuperscript{+} ions. The measurement uncertainty represents the day-to-day variation of the measured plasma composition, however the actual error in the plasma composition due to the internal settings of the mass spectrometer may be higher\cite{Schiesko2008}. 

\subsection{\normalsize{Temperature controlled sample holder}}\label{sec:sampleholder}

\begin{figure*}
	\centering
		\includegraphics[]{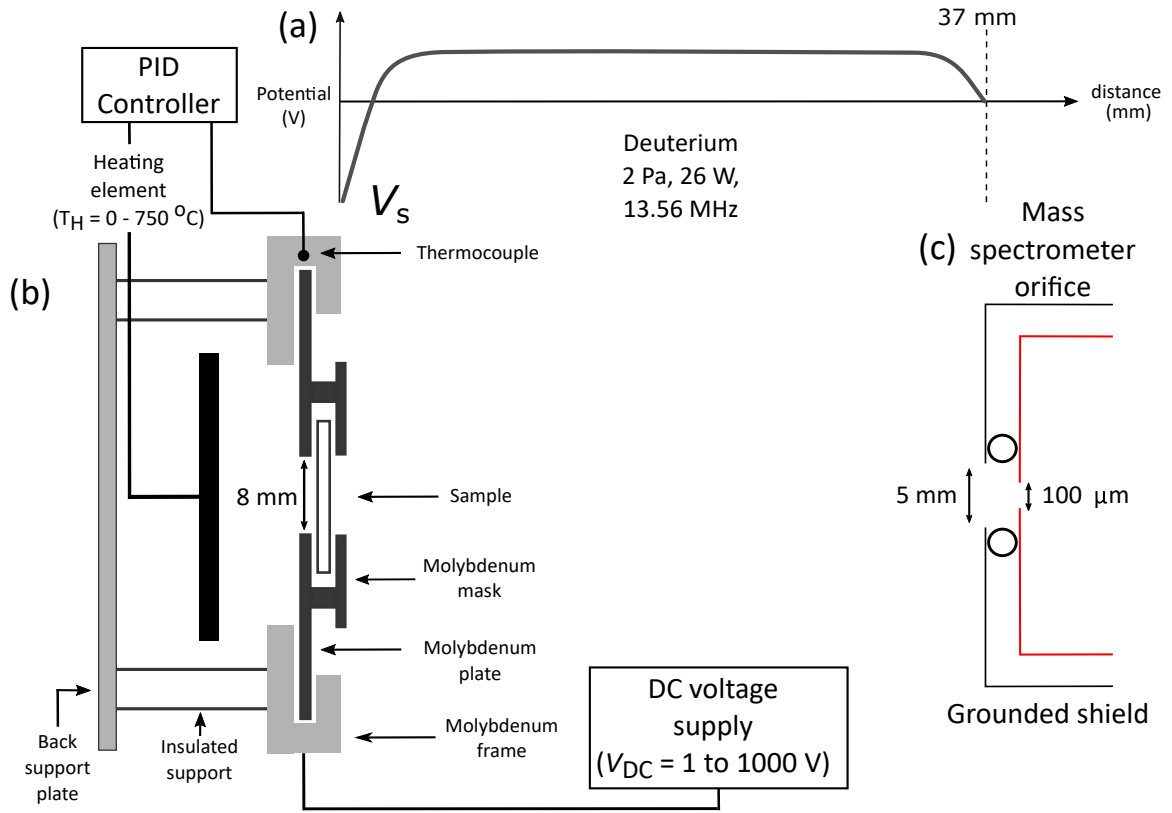}
		\caption{Schematic of the experimental setup. (a) Representative plasma parameters and potential profile between the biased sample surface at \textit{V\textsubscript{S}} and the mass spectrometer, 37~mm away. (b) Sample holder with heating element and thermocouple. (c) Mass spectrometer showing the external grounded shield and internal extractor orifice.}
    \label{Experiment}
\end{figure*}

The sample holder is shown in figure~\ref{Experiment}~(b). It is attached to a DC voltage source (Equipment Scientific Alimentation de Laboratoire CN7C) that can negatively bias the frame that holds the diamond sample. The voltage applied to the sample is defined as \textit{V\textsubscript{DC}}, which is distinct from the voltage at the sample surface, \textit{V\textsubscript{S}}. The sample is positioned 37~mm away from, and perpendicular to, the plane of the mass spectrometer orifice. This is the closest distance that the sample can be placed in front of the mass spectrometer orifice, and is assumed to be sufficiently small to achieve minimal negative ion signal loss. It has previously been demonstrated that this distance has negligible effect on the shape of the negative ion distributions measured by the mass spectrometer \cite{Ahmad2014, Kogut2019}.

It is worth noting that the angular dependence of the NIEDFs for carbon materials has previously been shown to be similar~\cite{Kogut2017a, Kogut2019, Dubois2016}. Therefore, a single measurement can be used to compare between samples. A misalignment of the sample surface normal to the mass spectrometer would produce spurious results, so to prevent this, the alignment is regularly checked by rotating the sample and maximising the negative ion signal.

As shown in figure~\ref{Experiment}~(b), a tungsten heating element is built into the sample holder, which is used to heat the back of the sample. The heating element is controlled by a PID controller (designed and built by AXESS tech) using a K-type thermocouple inside the the frame of the sample holder. 
By fixing a second thermocouple to the surface of the sample, its temperature is calibrated against the temperature measured by the PID. The heating element behind the sample increases the temperature of the sample's surface up to (750$\pm$20)$^{\circ}$C.

\subsection{\normalsize{Mass spectrometry for the measurement of negative ions}}\label{sec:Measurementandmassspec}

\subsubsection{\small{Mass spectrometer setup}}\label{sec:massspec}
An electrostatic quadrupole plasma mass spectrometer with attached energy analyser (Hiden EQP 300) is positioned in front of the sample surface as shown in figure~\ref{Experiment}~(c). The mass spectrometer has a \SI{100}{\micro\metre} diameter polarisable orifice separated from the main chamber by a 5~mm hole in a grounded shield. A grounded screen is positioned above the mass spectrometer orifice to reduce radio-freuency (RF) fluctuations from the plasma source (not shown in figure~\ref{Experiment}) \cite{AAhmad12013, Kogut2017a}. 

The mass spectrometer polariseable orifice potential is calibrated so that a nearly planar plasma sheath is formed in front of the orifice, as determined by a particle-in-cell (PIC) simulation \cite{AAhmad12013}. 

The potential on the surface of the samples accelerates any negative ions created through surface interactions away from the sample and through the plasma to the mass spectrometer. The low pressure of the plasma means there are few collisions between the plasma and the negative ions~\cite{Schiesko2008, AAhmad12013}. Any collisions that do occur with the deuterium plasma would predominantly be detachment collisions with deuterium molecules which would neutralise the negative ions,  thus preventing measurement of negative ions that have undergone collisions\cite{Hemsworth2005, Schiesko2008, AAhmad12013}. The plasma potential in front of the mass spectrometer prevents negative ions generated through volume production processes in the plasma from entering the mass spectrometer, therefore the energy of any negative ions that are measured must have been accelerated away from the sample surface\cite{Schiesko2008, AAhmad12013}. The negative ions are detected at an energy corresponding to the energy they possessed when they were created which can then be shifted by the kinetic energy gained between the sample and the mass spectrometer\cite{Schiesko2008, AAhmad12013}. Presented NIEDFs are shifted to present the kinetic energy the negative ions have at the surface of the samples. The secondary electrons emitted from the surface of the sample are filtered out within the mass spectrometer. 

Positive ions impacting the samples are assumed to dissociate during impact \cite{Cartry12017, Babkina2005a}, splitting the energy of the ion into its component particles (ie. for D\textsubscript{3}\textsuperscript{+}, 3 deuterium nuclei). This means that because the plasma is predominantly composed of D\textsubscript{3}\textsuperscript{+} ions, the modal energy of the ions striking the samples' surface is \textit{E\textsubscript{M}~=~e(V\textsubscript{S}+V\textsubscript{p})/3}, where \textit{V\textsubscript{p}} is the plasma potential, giving approximately 11~eV~per~particle at \textit{V\textsubscript{S}}~=~$-20$~V and 48~eV at \textit{V\textsubscript{S}}~=~$-130$~V. We define these conditions as `low energy' ion bombardment and `high energy' ion bombardment, respectively.  

The choice of $-130$~V is made to align with previously published work, whilst $-20$~V is chosen as this is the lower limit of what can be reasonably used to ensure effective self-extraction of negative ions from the sample surface into the mass spectrometer \cite{Ahmad2014, Kogut2019}.

\subsubsection{\small{Procedure for measurement of negative ion energy distribution functions (NIEDFs)}}\label{sec:measurement}

Measurements were undertaken using the following method:

\begin{itemize}
    \item The plasma was brought to steady state as determined by measurements of the positive ion energy distributions using the mass spectrometer
    \item A bias of either \textit{V\textsubscript{DC}}~=~$-20$~V or \textit{V\textsubscript{DC}}~=~$-130$~V was applied to the sample
    \item Negative ions produced following positive ion bombardment accelerate through \textit{V\textsubscript{S}}, cross the plasma volume and enter the mass spectrometer where the NIEDF was measured for sample temperatures between 30$^{\circ}$C and 750$^{\circ}$C in increments of 50$^{\circ}$C
    \item In order to compare the negative ion production yields for distinct material samples, the positive ion current was measured to the sample surface at 30$^{\circ}$C for \textit{V\textsubscript{DC}}~=~$-20$~V and \textit{V\textsubscript{DC}}~=~$-130$~V, using a copper electrode in the place of a sample which was insulated from the sample frame\cite{Kogut2019}.This method of measurement could not be used at high temperatures due to a temperature sensitive insulator used to isolate the copper electrode from the sample holder frame. Instead, in order to roughly monitor changes in the positive ion flux onto the sample, the positive ion current to the entire sample holder was measured using an ammeter connected to the frame of the sample holder. This showed that there was a thermal drift in the positive ion current to the entire sample holder of approximately 5\% irrespective of sample at both $-20$~V and~$-130$~V applied biases.
    \item The negative ion counts for each sample were integrated with respect to energy and then divided by the positive ion current measured neglecting the possible small changes with temperature to the isolated sample to give the relative negative ion yield for the sample\cite{Kogut2019}. This is given in arbitrary units as the mass spectrometer is not calibrated to count an absolute number of negative ions.
\end{itemize}

Recent measurements show that an absolute negative ion flux could be measured using a magnetised retarding field energy analyser via the technique described in Ref.~\citenum{Rafalskyi2015c}. A detailed investigation of this topic remains the subject of future work, but to provide some context for the presented results, preliminary measurements have shown that the yield from HOPG at 30$^{\circ}$C and a bias of~$-130$~V was approximately 1\%~\cite{G.}, compared to caesium , for which previous studies have reported yields of 30\%~\cite{Bacal2012}.

\subsection{\normalsize{Sample preparation}}\label{sec:Sampleprep}

\subsubsection{\small{Micro-crystalline boron doped diamond and micro-crystalline diamond}}\label{sec:MCDMCBDD}

As described in detail in Ref. \citenum{Ahmad2014}, non-doped and boron doped micro-crystlline diamond films, MCD and MCBDD respectively, were prepared in a bell jar reactor using plasma enhanced chemical vapour deposition (PECVD). The boron doped samples used in this study are comparable to the samples used in previous works where the gas phase doping level used is high (1000~ppm) to ensure a fully conductive diamond layer. The method of the creation of MCD and MCBDD samples is described elsewhere \cite{Ahmad2014}.

\subsubsection{\small{Micro-crystalline nitrogen doped diamond}}\label{sec:NDD}

The nitrogen doped diamond films were created using a similar PECVD technique to the MCD and MCBDD samples of Ref.\citenum{Ahmad2014} so only a brief summary is provided here. The PECVD process utilised a bell jar reactor with a pressure of 200~mbar, microwave power at 3~kW, substrate temperature of 850$^{\circ}$C, background hydrogen gas mixture with a methane concentration of 5\%. The ratio of nitrogen in the gas mixture was set as a means to vary the concentration of nitrogen in the MCNDD film. Each film was deposited on to a $($100$)$ orientated silicon wafer.

\subsection{\normalsize{Surface characterisation}}\label{sec:surfacestudy}

The samples were analysed using confocal microscopy and Raman spectroscopy prior to plasma exposure to characterise their properties.

\subsubsection{\small{Surface morphology and crystal structure}}
\begin{figure*}[t]
    \centering
    \includegraphics[]{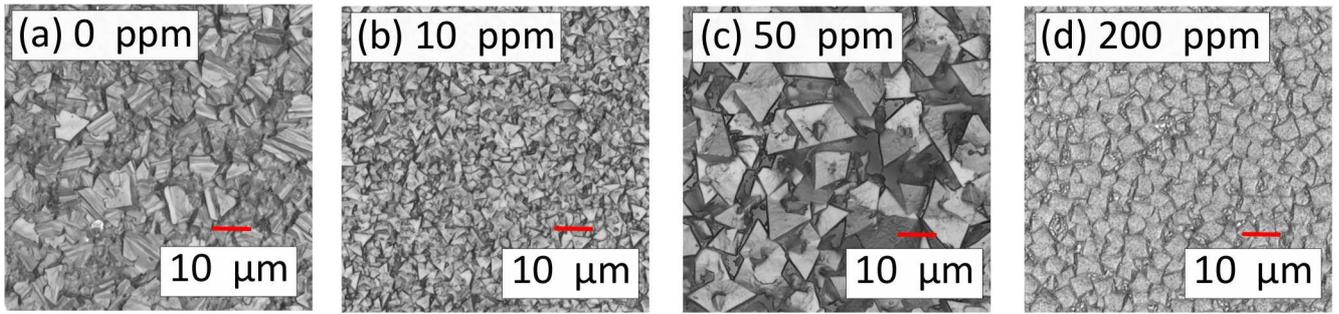}
    \caption{Confocal microscopy images of the diamond films grown with gas phase concentrations of (a) 0~ppm (b) 10~ppm (c) 50~ppm and (d) 200~ppm. The diamond crystals change in shape and size between the 4 gas phase dopant concentrations, from a mixture of $($111$)$ and $($100$)$ faces at 0~ppm and 10~ppm to predominantly $($111$)$ faces at 50~ppm to predominantly $($100$)$ crystal faces at 200~ppm.}
    \label{fig:Confocal_comp}
\end{figure*}

A laser confocal microscope (S neox, Sensofar) was used to observe the diamond surface morphology as shown in figure~\ref{fig:Confocal_comp}. From visual inspection, the crystal grains are observed to have grown to exhibit $($111$)$ crystal faces, $($100$)$ crystal faces, or a mixture of both, dependent on the concentration of nitrogen dopant introduced in the gas phase during sample growth \cite{Silva1998a}. As the gas phase nitrogen concentration is increased from 0~ppm to 50~ppm the crystals are observed to exhibit an increased proportion of $($111$)$ faces, with predominantly $($111$)$ faces observed at 50~ppm. As distinct from these, the crystal grains of the diamond film with 200~ppm gas phase doping displays predominantly $($100$)$ faces. The diamond films grown with 200~ppm nitrogen concentration in the gas phase are different to the other samples, due to be large crystals interspersed with regions of what appears to be much smaller crystals with a less pronounced crystal orientation. The size of the crystals for the 0~ppm sample, figure~\ref{fig:Confocal_comp}~(a), are approximately \SI{10}{\micro\metre}, whilst the 10~ppm sample, figure~\ref{fig:Confocal_comp}~(b), has a much smaller average crystal size, at approximately \SI{1}{\micro\metre}. The 50~ppm sample, figure~\ref{fig:Confocal_comp}~(c), has a crystal size similar to the 0~ppm sample, at approximately \SI{10}{\micro\metre}. The 200~ppm sample, figure~\ref{fig:Confocal_comp}~(d), as previously described, appears to have a distribution of large crystals separated by smaller crystals, here the average crystal size of the larger crystals is approximately \SI{5}{\micro\metre}.

\subsubsection{\small{Measurement of the relative quantity of nitrogen doping within the films}}\label{sec:NDDconcentration}

\begin{figure}[H]
    \centering
    \includegraphics[]{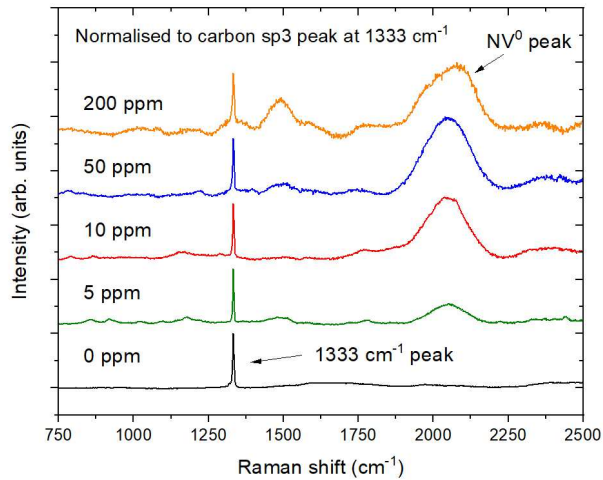}
    \caption{Raman spectra of nitrogen doped diamond samples taken at the centre of a dominant crystal face. The spectra are presented with the background subtracted (in order to aid clarity and comparison between samples) and normalised to the carbon sp3 peak, observed here at 1333~cm\textsuperscript{-1}.  The laser wavelength used for the measurements is 514~nm.}
    \label{fig:Raman_preplasma}
\end{figure}

Raman spectroscopy was undertaken to measure the relative concentration of nitrogen dopant that is introduced into the MCD films with respect to the gas phase nitrogen concentration present during the PECVD process. Raman spectra were generated using a Horiba Jobin Yvon HR800 setup. The measurements were undertaken in air using an excitation wavelength of $\lambda$\textsubscript{L}~=~514~nm, $\times$100~objective (numerical aperture of 0.9, i.e.~theoretical spot radius of \SI{0.34~}{\micro\meter}), 600 grooves/mm grating (resulting in a resolution of about 1~cm\textsuperscript{-1}), and 5~mW laser power. Measurements were taken using 5 acquisitions of 1 second intervals over a range of 0~cm\textsuperscript{-1} to 3000~ cm\textsuperscript{-1}, of which 750~cm\textsuperscript{-1} to 2500~cm\textsuperscript{-1} is presented,  with the measurement taken at the centre of a dominant crystal near the centre of the sample. A brief optical microscopic inspection of the samples prior to taking Raman measurements showed relatively uniform crystal distribution across the surface of all the samples and allowed for precise targeting of a dominant crystal near to the centre of the sample to be the subject of the Raman measurement. A quantitative comparison between Raman spectra has not been carried out as this requires the measurement of the polarisation of the Raman emission and a good understanding of the grain orientation~\cite{Mossbrucker1996}.

Figure~\ref{fig:Raman_preplasma} shows the Raman spectra from samples of nitrogen doped diamond with gas phase nitrogen doping of 0~ppm to 200~ppm. The spectra have been presented with the background fluorescence removed and normalised to the 1333~cm\textsuperscript{-1} peak, which can be attributed to sp3 bonded carbon (the diamond bond of carbon) \cite{Prawer2004, Tachibana2001}. The normalisation to this peak is justified due to the transparency of the diamond films such that the measurement is integrated across the sample thickness. The normalised spectra can therefore enable a comparison between samples that accounts for any change in the thickness of the MCNDD film \cite{Prawer2004}. 

The broad peak centred at 2100~cm\textsuperscript{-1} observed in figure~\ref{fig:Raman_preplasma} can be attributed to nitrogen vacancy centres (NV\textsuperscript{0}) that have been introduced into the diamond \cite{Kumar2016}. This broad peak appears, not due to vibrational modes, but due to the electronic signature attributed to nitrogen vacancy centres and in reality lies at an energy level of 2.15~eV. As Stokes Raman spectroscopy is energy loss spectroscopy, this peak appears arbitrarily at 2100~cm\textsuperscript{-1} when using a \SI{514}{\nano\meter} laser. Using another laser to perform the Raman spectroscopy results in a change in the wavenumber of this peak\cite{SmithRaman}. 

As the measurement configuration is the same for all samples, a relative comparison of the number of nitrogen centres in the diamond can be made using the broad 2100~cm\textsuperscript{-1} peak. This can then be used to infer relative nitrogen concentration \cite{Achard2007a}. As shown in figure~\ref{fig:Raman_preplasma}, the ratio of the NV\textsuperscript{0} peak to the peak centred at 1333~cm\textsuperscript{-1} increases with increasing gas phase dopant concentration, for samples 0~ppm to 50~ppm (200~ppm will be discussed below). This is consistent with previous work, which showed a similar increase in the magnitude of the NV\textsuperscript{0} characteristic peak with an increase in the gas phase nitrogen doping \cite{Tallaire2006}. 

In figure~\ref{fig:Raman_preplasma}, the Raman spectrum of the 200~ppm nitrogen doped diamond has a peak at 1500~cm\textsuperscript{-1} that has a much higher intensity than the other samples. This peak is of particular interest as it is associated with the sp2 bond of carbon that has previously been associated to graphite-like bonds \cite{Prawer2004}. The ratio of the peaks at 1333~cm\textsuperscript{-1} and 1500~cm\textsuperscript{-1} implies that there is a higher ratio of graphite in the 200~ppm diamond film compared to the other samples \cite{Chu2006, Ahmad2014, Merlen2017}. The 200~ppm nitrogen doped diamond sample also exhibits a NV\textsuperscript{0} centre peak at 2100~cm\textsuperscript{-1}, which is slightly lower than the 50~ppm sample, suggesting a reduction in the number of nitrogen vacancies, and therefore, a reduction in the concentration of nitrogen in the diamond. 

The observed increase in intensity of the NV\textsuperscript{0} peaks, increasing from 0~ppm to 50~ppm, may be attributed to both the increase in nitrogen introduced in the gas phase and by a change in the crystal face from a mix of $($100$)$ and $($111$)$, figure~\ref{fig:Confocal_comp}(a)~and~(b), to a primarily $($111$)$ face for which impurity incorporation is higher than that for $($100$)$ crystals, figure~\ref{fig:Confocal_comp}~(c).   This is consistent with the results of previous work \cite{Tallaire2006, Samlenski1995, Satoh1990}  This same process may then account for the slightly lower 2100~cm\textsuperscript{-1} peak for the 200~ppm sample compared to the 50~ppm samples despite a four fold increase in the nitrogen gas phase content. This decrease could be attributed to the change in the crystal orientation (see figure~\ref{fig:Confocal_comp}~(c)~and~(d)), from a $($111$)$ dominant crystal surface for the 50~ppm sample to predominantly $($100$)$ crystal orientation for the 200~ppm sample.

The surface characterisation of the samples show that the incorporation of nitrogen into the PECVD process has multiple effects on the diamond produced, aside from only substitutional or interstitial incorporation of nitrogen into the diamond lattice. For these samples, separating the difference in negative ion yield due to the influence of the crystal face or the nitrogen content in the diamond is not possible because of the interrelated nature the presence of nitrogen in the gas phase has with the crystal face orientation and the measurable number of nitrogen vacancy centres. This is an active area of research \cite{Zhao2019}. However for this study, as nitrogen doping is the main influencing factor that generates the differences between the samples, it is reasonable to suggest that it is possible to associate the nitrogen gas phase doping with the negative ion yield and this is how the samples will be defined in the next section. 

\section{Results}\label{sec:results}

\subsection{\normalsize{Nitrogen doped diamond: influence of the dopant concentration}}\label{sec:dopingconcentration}

\begin{figure*}
	\centering
		\includegraphics[]{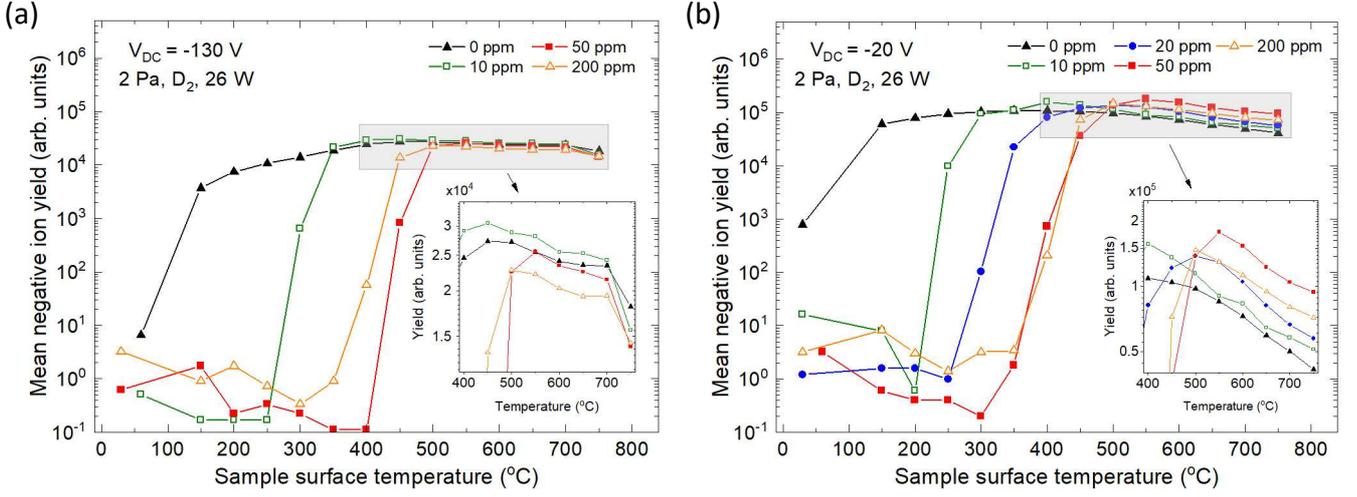}
		\caption{Negative ion yield plotted with respect to sample temperature for micro-crystalline nitrogen doped diamond (MCNDD) of doping concentration between 0~ppm and 200~ppm for (a) \textit{V\textsubscript{DC}}~=~$-130$~V and (b) \textit{V\textsubscript{DC}}~=~$-20$~V. Low pressure deuterium plasma operated at 2~Pa and 26~W. Insets for both (a) and (b) depict the highlighted region's negative ion yield for temperatures between 400$^{\circ}$C and 750$^{\circ}$C. Solid lines have been added to guide the eye.}
    \label{fig:_MCNDD_doping_comparison}
\end{figure*}

Figure~\ref{fig:_MCNDD_doping_comparison} presents the negative ion yield from MCNDD for different dopant concentrations, as measured in the gas phase during sample preparation. In both figures~\ref{fig:_MCNDD_doping_comparison}~(a)~and~(b) the yield profile for MCNDD has a distinct shape. For example, at 50~ppm, the measured yield is practically zero between a temperature of  30$^{\circ}$C and 400$^{\circ}$C. At 450$^{\circ}$C, the yield rapidly increases by several orders of magnitude to a maximum at 550$^{\circ}$C. This transition is similar for all nitrogen doped samples with the transition occurring at temperatures ranging from 250$^{\circ}$C to 450$^{\circ}$C. This is somewhat unlike MCD which produces measurable negative ions for all temperatures. The trend for MCD is a gradual increase to a maximum yield at a temperature of 450$^{\circ}$C ($-130$~V, high energy bombardment) or 400$^{\circ}$C ($-20$~V, low energy bombardment), there is then a decrease in yield from this maximum yield as the temperature is increased further to the maximum temperature of 750$^{\circ}$C. It can also be observed that between $\sim$30$^{\circ}$C and $\sim$150$^{\circ}$C, MCD does undergo a transition, though this is smaller than that seen for MCNDD. 

In order to understand these trends, it is important to note that the negative ion yield measured using the technique described in this article relies on a conductive sample surface. A non-conductive sample would not allow negative ions to be accelerated to the mass spectrometer at an energy which the mass spectrometer is tuned for. The magnitude of the transitions seen in figure~\ref{fig:_MCNDD_doping_comparison}
is a feature of this experimental technique which inadvertently highlights the temperature at which samples become conductive, such that  \textit{V\textsubscript{DC}}~=~\textit{V\textsubscript{S}}. 

The difference in trends between MCNDD samples and the un-doped diamond can be attributed to the differences in conductivity between MCNDD and MCD. Previous work with MCD has shown that it has poor conductivity close to room temperature, which explains the increase in yield occurring between 30$^{\circ}$C and 150$^{\circ}$C \cite{Cartry12017}. Regarding MCNDD, the temperature at which the sharp increase in yield occurs appears to be dependent on the nitrogen doping of the diamond sample. The results of previous work suggest that the level of interstitial nitrogen doping influences the conductivity of diamond \cite{Baranauskas1999, Al-Riyami2010, Bhattacharyya2001} and this sharp increase is consistent with increasing nitrogen dopant concentration and its influence on the conductivity of the diamond, supporting the argument that the nitrogen incorporated into the diamond increases as gas phase nitrogen is increased during its production. An exception to this trend are the results for 200~ppm in both figure~\ref{fig:_MCNDD_doping_comparison}~(a)~and~(b), which does not exhibit a significant increase in the sample temperature for which the film becomes conductive relative to the 50 ppm MCNDD sample. This can be explained by considering figure~\ref{fig:Raman_preplasma}. As discussed in section~\ref{sec:NDDconcentration}, the nitrogen content measured using Raman spectroscopy suggests a nitrogen content that is similar for 200~ppm and 50~ppm MCNDD samples meaning, in the absence of other influences, a similar conductivity for these two samples could reasonably be expected. 

The maximum yield from each sample occurs at temperatures between 400$^{\circ}$C and 550$^{\circ}$C, which is highlighted in the insets of figures~\ref{fig:_MCNDD_doping_comparison}~(a)~and~(b). For the 0~ppm and 10~ppm samples, the maximum yield occurs at 400$^{\circ}$C, whilst for 20~ppm (and 200~ppm) it occurs at 500$^{\circ}$C and for 50~ppm, at 550$^{\circ}$C. As mentioned previously, a conductive sample surface is necessary to hold a DC surface bias which is necessary for the acceleration of negative ions into the mass spectrometer. The trend of increasing temperature for maximum yield as dopant increases (excluding 200~ppm) could be related to the maximum yield in these experimental conditions being restricted by the conductivity of the samples. For example, the maximum yield for MCD and MCNDD~(10~ppm) is at the peak of a gradual increase and decrease in yield as temperature is increased from $\sim$30$^{\circ}$C to $\sim$400$^{\circ}$C and then from $\sim$400$^{\circ}$C to $\sim$750$^{\circ}$C respectively. This is most clearly observed in figure~\ref{fig:_MCNDD_doping_comparison}~(b) for the 10 ppm sample. This sample is distinct from the other MCNDD samples as the 10 ppm sample exhibits an increase in yield as the temperature is increased (due to a change in conductivity) then a further smaller increase up to a maximum negative ion yield at$\sim$400$^{\circ}$C. The yield then gradually decreases as the temperature is increased further. The other MCNDD samples also undergo an increase in yield due to a change in conductivity, but no further gradual increase in yield is observed as temperature is increased.

It could therefore be reasonable to suggest that the peak yield conditions are not observed due to a lack of conductivity for samples with more than 20~ppm gas phase nitrogen doping. The trends observed for the 0~ppm and 10~ppm samples suggest that a temperature of approximately 400$^{\circ}$C may be the temperature at which these MCNDD with more than 20~ppm gas phase doping produce the highest yield. A technique to measure negative ions that does not require a conductive surface would be necessary to explore this further.

In figure~\ref{fig:_MCNDD_doping_comparison}~(a), the effect of the nitrogen doping on the maximum yield is not readily observed when a bias voltage \textit{V\textsubscript{DC}}~=~$-130$~V is applied to the sample. This is unlike figure~4~(b) in which a bias voltage of \textit{V\textsubscript{DC}}~=~$-20$~V is used. In this data there is an observed difference between the nitrogen doped and non-doped diamond. The yield in figure~4~(a) for the MCNDD samples and MCD samples is also lower than the yields from all of the samples in figure~4~(b). A higher bombardment energy as a result of the high magnitude bias is associated with an increase in sp2 bond formation in diamond \cite{Dubois2016}. It is reasonable to suggest that the reduction in yield for the higher magnitude bias creates more sp2 defects which decreases the yield. Additionally, if the yield is not changing with the addition of nitrogen to the diamond, it is also reasonable to suggest that the nitrogen doped diamond may be more susceptible to defect formation due to high energy bombardment which would result in a surface state that does not enhance the negative ion yield through nitrogen doping. Additional work would be necessary to characterise this process.

The apparent influence of nitrogen doping on the measured negative ion yield is observed in figure~\ref{fig:_MCNDD_doping_comparison}~(b) where a sample bias voltage \textit{V\textsubscript{DC}}~=~$-20$~V is applied. When comparing the yield at temperatures above 550$^{\circ}$C, i.e. when all of the MCNDD films are conductive, it is observed that the negative ion yield is higher at similar temperatures, when increasing nitrogen dopant concentration for the 0~ppm to 50~ppm cases. The mechanism for such an increase in yield is not immediately clear and future work will be necessary to identify the specific cause of this increase. For example, it could be solely due to interstitial nitrogen, or a change of crystal orientation or a combination of both. In any case, the increase is correlated to the amount of nitrogen dopant with the exception of the result observed for 200~ppm gas~phase doping, which produces a comparatively lower yield compared to the 50~ppm case. Should interstitial nitrogen content be the main cause of an increase in negative ion yield, this result can be explained by the Raman measurements shown in figure~\ref{fig:Raman_preplasma}. As discussed in section~\ref{sec:NDDconcentration}, the Raman measurement suggests that the diamond has a similar amount of nitrogen doping when comparing the 2100~cm\textsuperscript{-1} peak for the 50~ppm and 200~ppm samples. However the Raman measurement also suggests that the 200~ppm MCNDD sample has more carbon sp2 bonds (graphite-like) than the 50~ppm sample. The reduction in yield observed for the MCNDD~(200~ppm) sample compared to MCNDD~(50~ppm) sample is therefore consistent with previous work, which observed that an increased number of sp2 bonds is less favourable to negative ion production\cite{Kogut2019, Ahmad2014}. This work suggests that this is still the case with nitrogen doped diamond samples. 

\subsection{\normalsize{Mechanism for the surface production of negative ions}}\label{sec:mechanisms}

\begin{figure*}
	\centering
		\includegraphics[]{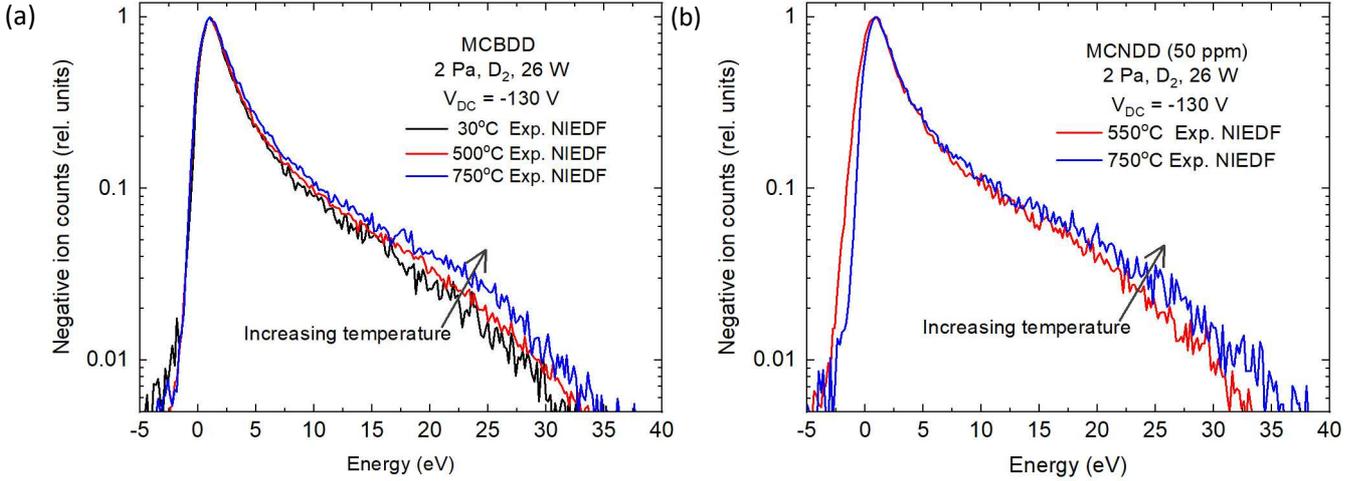}
		\caption{NIEDFs for: (a) micro-crystalline boron doped diamond (MCBDD) at 30$^{\circ}$C, 500$^{\circ}$C and 750$^{\circ}$C, (b) micro-crystalline nitrogen doped diamond (MCNDD) at 550$^{\circ}$C and 750$^{\circ}$C. Increases in sample temperature lead to a decrease in the number of low energy negative ions, which results in an increase in the height of the tail of high energy negative ions when the distribution is normalised. Low pressure deuterium plasma operated at 2~Pa and 26~W.}
    \label{fig:_MCBDD_NIEDF}
\end{figure*}

NIEDFs for MCBDD and MCNDD are presented in figure~\ref{fig:_MCBDD_NIEDF} to compare negative ion production processes between MCBDD, a previously studied material \cite{Cartry12017}, and the MCNDD samples. In this figure the NIEDFs are normalised to the modal negative ion energy at temperatures where MCBDD and MCNDD samples are conductive.

When considering a normalised NIEDF, a reduction in the proportional magnitude of the NIEDF peak at low energies will result in an increase in the apparent proportion of negative ions at high energies. The NIEDFs in figure\ref{fig:_MCBDD_NIEDF}~(a), show that for MCBDD the proportion of high energy ions increases as the surface temperature increases.
This is because the main contribution to the measured yield is low energy ions, which are predominantly created through the sputtering process, as distinct from backscattering, due to the acceptance angle of the mass spectrometer \cite{AAhmad12013, Cartry12017}. Previous work has confirmed this interpretation through comparison of experimental results with SRIM simulations \cite{Dubois2016}. 
The physical interpretation for the decrease in the sputtering contribution is that this is due to a decrease in the amount of sub-surface deuterium available for sputtering as a result of out-gassing caused by the increase in temperature. \cite{Dubois2016, Kogut2017a}. 

At a surface bias of \textit{V\textsubscript{S}}~=~$-20$~V, i.e. the `low energy' bombardment condition described in section~\ref{sec:Measurementandmassspec}, the high energy tail observed in the NIEDFs in figure~\ref{fig:_MCBDD_NIEDF} is not produced. Without a high energy tail, the normalised NIEDF shapes are not strongly dependent on the deuterium surface content \cite{Kogut2019}, and comparison of the ratio of sputtered to backscattered particles cannot be readily inferred using this approach. For this reason, only results with a surface bias of \textit{V\textsubscript{S}}~=~$-130$~V are presented.

Comparing figure~\ref{fig:_MCBDD_NIEDF}~(a) to figure~\ref{fig:_MCBDD_NIEDF}~(b), which presents NIEDFs for MCNDD at 550$^{\circ}$C and 750$^{\circ}$C, i.e. temperatures at which the sample is conductive, it is observed that MCNDD displays a similar increase in the proportion of high energy negative ions as the sample's surface temperature increases. This implies that MCNDD has similar negative ion production properties to MCBDD.
 
The trends for MCNDD and MCD observed in figure~\ref{fig:_MCNDD_doping_comparison} and discussed in the previous section can be explored in the context of figure~\ref{fig:_MCBDD_NIEDF}. Figure~\ref{fig:_MCNDD_doping_comparison} shows that the yield for MCD increases up to sample temperatures of $\sim$400$^{\circ}$C and decreases as its temperature is increased further. This is similar to the trends observed for samples of MCNDD when they are conductive. The increase and then decrease in yield as temperature is increased, from $\sim$30$^{\circ}$C to $\sim$400$^{\circ}$C and then from $\sim$400$^{\circ}$C to $\sim$750$^{\circ}$C respectively, can be attributed to two processes that combine to generate the observed trend in figure~\ref{fig:_MCNDD_doping_comparison}. The first process is the removal of defects on the sample surface. The heating of the sample results in an enhancement of the etching of sp2 bonds created by the bombarding positive ions resulting in a surface which results in a higher ratio of sp3 bonds~\cite{Ahmad2014}. The increased proportion of diamond bonds on the surface increases the negative ion yield, as explored in previous work through Raman spectroscopy~\cite{Ahmad2014, Cartry12017, Kumar2011, Kogut2019}. The second process is the previously discussed decrease in the sputtering contribution to the negative ion yield due to out-gassing of deuterium from the sample surface, as observed in the measurements of figure~\ref{fig:_MCBDD_NIEDF}. As temperature is increased, the influence of each of these processes on the measured negative ion yield is observed to vary significantly. At temperatures below $\sim$400$^{\circ}$C, the reduction in defects increases the yield, whilst the outgassing does not cause a significant decrease in the sputtering contribution. At temperatures above $\sim$400$^{\circ}$C, the decrease in sputtering contribution reduces the yield by a greater extent than the reduction in defects caused by the elevated temperature, causing a reduction in the the measured negative ion yield~\cite{Ahmad2014}.

For the samples of nitrogen doped diamond with more than 20~ppm nitrogen added in the gas phase, the MCNDD film is not conductive at temperatures where the previously mentioned reduction in the defects can increase yield, i.e. between 30$^{\circ}$C and 400$^{\circ}$C. A more thorough exploration of the resulting interplay between the reduction of defects and the decreasing sputtering contribution is beyond the scope of this experimental study. However, figure~\ref{fig:_MCBDD_NIEDF}~(b) suggests that the decrease in yield due to a decrease in the sputtering contribution is consistent with current understanding of the behaviour of negative ion formation on micro-crystalline doped diamond. 

\subsection{\normalsize{Negative ion yield: comparison between MCNDD, MCBDD, and MCD}}\label{sec:MCDMCBDDMCNDDres}

\begin{figure*}
	\centering
		\includegraphics[]{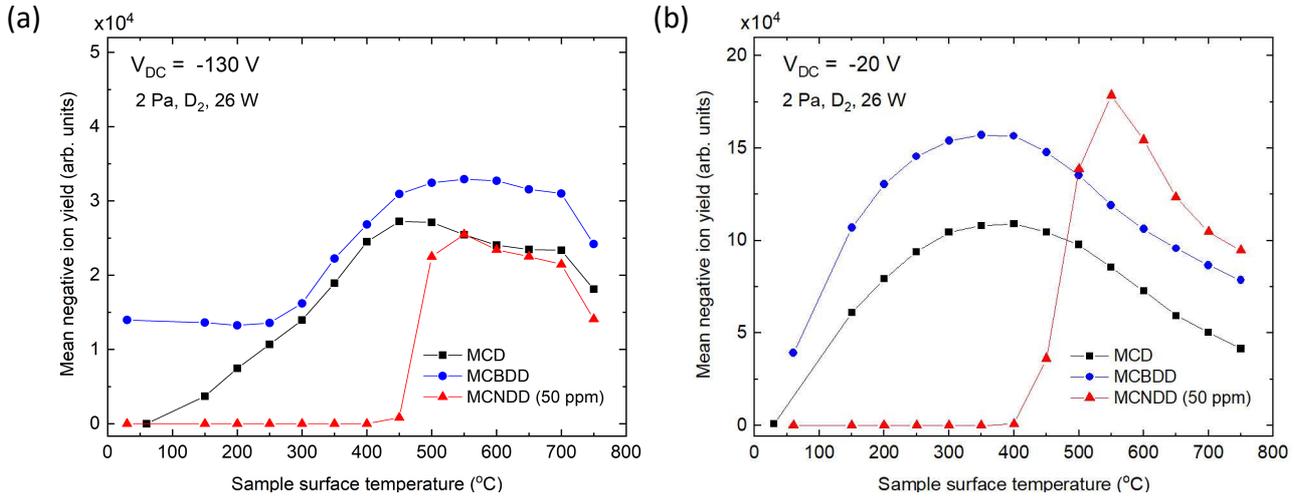}
		\caption{Negative ion yield with respect to film surface temperature for micro-crystalline diamond (MCD), micro-crystalline boron doped diamond (MCBDD) and micro-crystalline nitrogen doped diamond (MCNDD) for (a) \textit{V\textsubscript{DC}}~=~$-130$~V and (b) \textit{V\textsubscript{DC}}~=~$-20$~V. Low pressure deuterium plasma operated at 2~Pa and 26~W. Solid lines have been added to guide the eye.}
    \label{MCDMCDBDDMCNDD}
\end{figure*}

The negative ion yield with respect to sample surface temperature of the MCD, MCBDD and MCNDD samples is shown in figure~\ref{MCDMCDBDDMCNDD}, with high energy ion bombardment (\textit{V\textsubscript{DC}}~=~$-130$~V) shown in figure~\ref{MCDMCDBDDMCNDD}~(a) and low energy bombardment (\textit{V\textsubscript{DC}}~=~$-20$~V) shown in figure~\ref{MCDMCDBDDMCNDD}~(b). 50~ppm MCNDD is chosen as a comparison to MCD and MCBDD as this produced the highest relative negative ion yield of all the nitrogen doped diamond samples, as shown in figure~\ref{fig:_MCNDD_doping_comparison}.

In figure~\ref{MCDMCDBDDMCNDD}~(a), \textit{V\textsubscript{DC}}~=~$-130$~V, the trends for MCD and MCBDD are similar, with an increase in yield by a factor of 6 from 150$^{\circ}$C to 450$^{\circ}$C observed for MCD and a factor of 2 observed for MCBDD from 150$^{\circ}$C to 550$^{\circ}$C. The yield then decreases gradually as temperature is increased further. These results are consistent with previous work using MCD and MCBDD\cite{Cartry12017}. The negative ion yield from MCNDD at sample temperatures below 400$^{\circ}$C is effectively zero. {After 400$^{\circ}$C} there is a rapid increase in yield by several orders of magnitude up to 550$^{\circ}$C, as discussed in section~\ref{sec:dopingconcentration}. After 550$^{\circ}$C, the trend agrees with MCD and MCBDD. In the high energy bombardment regime, the yield from MCNDD is found to be lower than MCBDD and comparable to MCD. This suggests that the higher positive ion bombardment energy is having a larger influence on MCNDD than MCBDD, though a mechanism for such a difference is beyond the scope of this study.

In figure~\ref{MCDMCDBDDMCNDD}~(b) for \textit{V\textsubscript{DC}}~=~$-20$~V, the trends for MCD and MCBDD are also observed to be qualitatively similar, showing an increase in yield by a factor of 2 and a factor 1.5 from 150$^{\circ}$C to 400$^{\circ}$C respectively, and a gradual decrease in yield above 400$^{\circ}$C, which has been discussed in section~\ref{sec:dopingconcentration} \cite{Cartry12017, Kogut2019}. Figure~\ref{MCDMCDBDDMCNDD}~(b) has a similar trend as figure~\ref{MCDMCDBDDMCNDD}~(a) where the yield from MCNDD at temperatures below 400$^{\circ}$C is effectively zero. The yield increases by several orders of magnitude between 400$^{\circ}$C to 550$^{\circ}$C, after which it decreases gradually. At temperatures above 550$^{\circ}$C the general trend of decreasing yield is consistent with both MCD and MCBDD, and agrees with current understanding of these diamond films as discussed in the previous section. Of particular interest is that the yield for MCNDD in this low energy ion bombardment condition is observed to be higher than MCD, and also higher than the previously best performing type of diamond, MCBDD\cite{Cartry12017}. At 550$^{\circ}$C, the maximum yield observed, MCNDD has a higher negative ion yield than MCD and MCBDD by a factor of 2 and 1.5, respectively. This therefore suggests that controlled addition of nitrogen during the growth of diamond using the PECVD process could be an avenue for increasing the negative ion yield from diamond.

\section{\large{Conclusion}}\label{sec:conclusion}

In this study, we have investigated the nitrogen doping of diamond films as a means of increasing the negative ion yield during exposure to a low pressure deuterium plasma (2~Pa, helicon source at 26~W). For conditions where positive ions from the plasma bulk bombard nitrogen doped diamond film with energies of 11~eV and 48~eV, `low energy' and `high energy' bombardment, respectively, mass spectrometry measurements are used to determine the negative ion yield as the film temperature is scanned between 30$^{\circ}$C and 750$^{\circ}$C. For 50~ppm nitrogen doping, introduced in the gas phase during diamond growth using the PECVD technique, the application of low energy ion bombardment is observed to increase the negative ion yield by a factor of 2 compared to un-doped diamond and a factor of 1.5 compared to boron doped diamond. 

\section{\large{Acknowledgements}}

The authors would like to acknowledge the experimental support of Jean Bernard Faure and the PIIM surface group. This work has been carried out within the framework of the French Federation for Magnetic Fusion Studies (FR-FCM) and of the EUROfusion consortium, and has received funding from the Euratom research and training programme 2014-2018 and 2019-2020 under grant agreement No.~633053. The views and opinions expressed herein do not necessarily reflect those of the European Commission. Financial support was received from the French Research Agency (ANR) under grant 13-BS09-0017 H INDEX TRIPLED. The financial support of the EPSRC Centre for Doctoral Training in fusion energy is gratefully acknowledged under financial code EP/L01663X/1. CGI (Commissariat \'a l'Investissement d'Avenir) is gratefully acknowledged for its financial support through Labex SEAM (Science and Engineering for Advanced Materials and devices) (No. ANR 11 LABX 086, IDEX 05 02). 

\section{\large{References}} 

\renewcommand\refname{}{\vskip -1cm}

\begin{footnotesize}
\bibliography{Bibliography}
\bibliographystyle{unsrtnat}
\end{footnotesize}


\end{document}